# Theoretical Calculation of Shrinking and Stretching in Non-Uniform Bond Structure of Monolayer Graphite Flake via Hole Doping Treatment

Osman Özsoy[1,2] and K. Harigaya[3]

[1]Electronics Technology, Kayseri Vocational Collage, Erciyes University, Kayseri, 38039 Turkey

[2]Erciyes Technopark Corporation, Erciyes University, Kayseri, 38039 Turkey

[3]Nanotechnology Research Institute, AIST, Tsukuba 305-8568 Japan

**Abstract:** This paper deals with the physics of monolayer graphite, with a particular focus on the electronics and structural properties. In contrast to the previous electronic band structure of doped single-walled carbon nanotube calculation [1], where just graphite plate is considered, here the carbon bonds length alteration is calculated in terms of hole doping. It is found that doped holes play crucial roles on the bond structure compared to that obtained from no doping configurations, and it changes as hole doping increases.

## Introduction:

A number of theoretical and experimental works for carbon-based materials have been produced over the years. One class of materials of increasing interest is one of the well-known graphitic materials which consist of densely packed carbon atoms arranged into a honeycomb lattice that is vital for life on the earth. Graphene has zero gap semiconductor material [2] in which the valence and conduction bands slightly overlap in energy. It is considered a promising candidate for future electronics and spintronic devices [3,4] that allow devices in tetrahertz order [5], transistors [4,6], novel sensors [7], semiconducting graphite oxide films [8], spin filters [9], graphene p-n junctions [10-13], p-n-p junctions [14]. It opens a avenue for all carbon-based microelectronics [15] while switching their function to atomic-scale in electronic devices [16]. In addition, for decades research has been concluded on graphene because of its high conductivity, anomalous quantum Hall effect, low scattering rate from defects and its electrons acting more like its electrons like photons than particles [16].

Monolayer graphene has been studied more frequently than bilayer graphene in the literature; the starting point for studying graphite was first done by Wallace [17]. Electronic compressibility [18], nonlinear elasticity [19], thermal conductivity and electron linewidth by using first-principles analysis [20,21], $\pi$-junction qubit [22], dynamics of wave packet [23], chiral tunneling via a time-periodic potential [24], quantum and transport conductivities [25], local density of states [26], spin filters [27], tremling motion of charge carriers [28], some electronics properties [29,30], ballistics transmission [31], phonon thermal conductance [32], carrier density and magnetism [33], some of the low energy electrical and magnetic properties [34], molecular dynamics study by using the Tersoff-Brenner potential [35], graphite nano-needle as cold cathodes [36] were some recent theoretical and experimental studies devoted to it.

On the other hand, doping/undoping against electronic and mechanical properties of graphene is an increasingly important field of study in nonoscale materials. The best method to alter those systems' properties is to dope/undope them since doping/undoping neither destroys the hexagonal graphene carbon network nor annihilates the electronic band structure. By doing this, it provides various possibilities for controlling the physical properties of the graphene. Various methods can be employed so doping ca be achieved with--nonmetals, alkali metals, transition metals, and clusters--the most commonly used form of doping known as exohedral doping [1,37].

In addition, some recent research studies of doped graphene include molecular doping [38], electron doping [39,40], hole doping [41], both S and P doping [42], the first principles method for p-type graphene [43], linear response to vector potentials [44], electronic structure and electron-phonon coupling [45], buckling [46]. The bond length modulation response against the doping/undoping treatment of graphene-based materials is also a particularly interesting topic [47-50].

Here we first consider the very small, pure graphene flake, and then we work with the tight-binding method [51] to explore the response of the monolayer neutral graphene flake to dope holes treatment since the contribution to bond length changes arising from the modulation of electron and hole doping by lattice distortions. Mutual distortions in the bond length of the honeycomb structure of graphene were compared with those of pure and undoped graphene cases.

**Method:**

Graphene originates from the carbon atom. Carbon is the sixth element of the periodic table (belongs to the group IV). It has six electrons (four electrons in the outermost shell are also called valance electrons), and its electron configuration is showed as $1s^2 2s^2 2p^2$ ($2p_x 2p_y$). The chemical bonds are created by those three valance electrons (form the $sp^2$ hybrid orbital planar trigonal symmetry) and undergo a $\sigma$ bond which is a covalent bond representing a strong bond--bonding which is extremely localized because of the large overlap of the integral. The remaining electron ($\pi$-electron) of the valance electron is in a $2p_z$ orbital which is oriented towards perpendicular to the graphene plane by sharing a pair of electrons. It leads the energy band structure of the graphene. Those delocalized π-electrons are necessary for π-bonds (weak bonds). In short, graphene is 2D material which has both σ and π-bond with $sp^2$ orbital.

Here, we carried out the tight-binding approximation and all formulations are the same as for Ref. [1] thus we did not give details here. The bond alternation is formulated as

$$v_i = 2\gamma\left[\frac{\alpha C}{\gamma} + \sum_{j,\sigma}' B_{i+1,j,\sigma}^{+} B_{i,j,\sigma}\right]. \tag{1}$$

In Eq.(1), **B**s are the eigenvectors, $\gamma = \frac{\alpha^2}{\kappa}$ where α is the electron-lattice displacement coupling constant taken as 6.31eV/Å, κ is the effective spring constant taken as 49.7eV/ $Å^2$ and then Lagrange multiplier C is found by adding up both sides of Eq. (1) with the condition $\sum v_i = 0$ as

$$C = -\frac{1}{N_b}\frac{\gamma}{\alpha}\sum_{j,\sigma}' B_{i+1,j,\sigma}^{\dagger} B_{i,j,\sigma}\ , \tag{2}$$

where $N_b$ is the total number of π-bonds, and the primes in the summations indicate the occupied states of electrons. Our calculation method yields the following steps: i) Hamiltonian matrix is calculated by arbitrary initial values of $v_i$'s. ii) By substituting Hamiltonian into the Schrödinger equation, the eigenvalues and then eigenvectors are found.

iii) Total energy and $v_i$'s are numerically calculated. iv) In the first and third steps, the given calculation was assumed to have converged once $v_i$ varied by less than $10^{-7}$.

We consider a very small size graphene flake for this study as seen in Fig. 1 since showing a large amount of atom interaction in graphics is limited. Nevertheless, the flake is so small since *width* x *length* = 6.93x5.81 nm. In the figure, labeled numbers represent carbon atoms, and the total carbon atoms are as few as 24. It would also be worthwhile to point out that 1st and 24th atoms have only one bond but the others have two and three. Thus one does not need to consider these two atoms' effect in the calculation since they are the most unstable.

Here we now study the contribution to the bond length changes arising from the modulation of hole doping into intrinsic graphene. As already given, $N_{\mathrm{b}}$ is the total number of π-bonds which is 29 (including two unstable bonds) for the flake. Number of holes $N_{\mathrm{p}}$ in the doping process is varied within $N_p \leq N$+10 or $0 \leq N_p \leq 0$ in which $N$ is the number of carbon atoms of graphene flake, and we calculate for all the possible $N_{\mathrm{p}}$ which reflects maximum 10 holes doping in this study.

Stretching and shrinking phenomena of the bonds in graphene are called bond distortion or relaxation in nm. The bond distortion of the graphene was analyzed as the number of holes increases, and the results were considered with and without doping holes. To gain physical insight into the problem, let us pay attention to a no doping case. Solid lines represent intrinsic graphene which is called no hole doping in all figures. The ordinate represents relaxation in nm in the graphs. Here, positive and negative relaxation means stretching and shrinking in the bond length, respectively. The abscissa also represents carbon atoms interactions in the 2D graphite flake. The bond between second and third C atoms seems to be shrinking since relaxation is negative, and it is about -0.00247 nm. The bond among atom 3. and 4. have small negative relaxation value of +0.00244 nm. As seen from the Fig.2(a), the bond between 7. and 8. atoms and 17. and 18. atoms have the same relaxation value of 0.0052 nm that has the maximum positive value. The most stable bond lines are between 12. and 13. C atoms. It is worthwhile to mention that formation of C-C bonds appear as shrinking and stretching in turn from atom 2 to 8 and from 17 to 23 while stretching and shrinking from 9 to 16 in turn. This bond relaxation is depicted in Fig.2(b). In the figure, (+) and (-) represent stretching and shrinking in the bond length changes, respectively. From the same figure, one can see that the length of the flake changes horizontally, but the width of the flake becomes smaller since interestingly all vertical bonds shrink. In short, maximum relaxations take place in outer bonds which are the edges of the flake. One can point out that the outer edges of the flake are more unstable than the inner bonds.

To understand the effect of hole doping on bond distortion, first we dope the flake with one hole, and its structure is compared as depicted in Fig. 2(a). From one hole doping to four holes doping can be called light doping. Solid lines represent intrinsic graphene which is called no hole doping in all figures. In addition, dashed and dotted lines represent one and two hole doping case, respectively.

The most stable bonds lead from 4 to12, 13 to 21 and 22 to 23 atoms while the most unstable bonds lead from 8 to 16 and 9 to 17 atoms in case of one hole doping. From the general stretching and shrinking case point of view, no big differences appear as compared to the pure flake. When the system is doped with two holes, all vertical bonds keep their former situation except the bonds between 7-8, 17-18 and 2-10 atoms which are the most unstable bonds with the value of $\pm0.0006$nm as seen in Fig.2(b).

The next possible doping step is three holes doping, and it is shown in Fig.3(a) and (b). All vertical bonds stretch out. There are three horizontal bonds lines, and the middle line containing atoms 9 through 16 shows shrinking. The most stable bonds are present between 7-8, 7-18, 4-12 and 13-21 atoms, and the bond distortions are in the order of magnitude +0.0000577 to 0.0000996 nm. The most unstable bonds consist of between 8-16 and 9-17 atoms with the value of +0.00672nm. In the case of four hole doping, 16 bonds out of 26 shrink and one can state that the flake is comparably smaller than the former one or the pure flake. The most stable bonds linked between 2-3 and 22-23 C atoms with the relaxation value of -0.001774nm.

Bond distortion gets its smaller value for the bonds between 2-3 and 22-23 atoms thus those are the most stable bonds as hole doping increases to five holes which is depicted in Fig. 4(a) and (b). Here the heavy doping corresponds to eight holes doping starting with five holes doping. The second horizontal bonds line behaves like a four holes doping case. When the flake is doped with six holes, the most stable bonds become between 3-4 and 21-22 atoms with the relaxation value of +0.000281nm which are in the neighborhood of the most stable bonds of a five holes doping case. This value is about ten times smaller than the former doping case.

Seven and eight holes doping cases in the bond distortion also show sinusoidal behavior as seen in Fig.5(a) and (b). The last part of the graph in the abscissa which meets for vertical bonds has similar behavior and gets the same relaxation values. Stretching bonds numbers which are 13 and 15 for seven and eight holes doping, respectively, becomes greater and greater with increasing holes doping. All vertical bonds almost stretch out, and the flake becomes wider in size.

Finally, nine and ten hole doping is depicted in Fig.6(a) and (b) and this case can be called the most heavy doping. Most of the bonds either vertically or horizontally stretch out, and the flake gets larger in both dimensions. 18 and 19 bonds out of 26 stretch for the nine and ten holes doping cases, respectively. Bonds between 6-14 and 11-19 C atoms are the most stable bonds for nine holes doping, while bonds 2-3 and 18-19 C atoms are the most stable for ten holes doping. Both graphs represent similar behavior in the last part of the abscissa which encounters vertical bonds of the flake. The most unstable bonds are between 7-8, 17-18, 2-10 and 15-23 C atoms, and their relaxation value is about ±0.0055nm

It may also be possible to infer from the average bond distortion of the whole system. Fig.7 shows average bond distortion of the graphene for three different types. Average distortion of tilt (horizontal), right (vertical) and mixed (all) bonds are depicted in the graph. The abscissa starts with zero which corresponds to the pure flake or no doping and end up with ten holes doping. In contrast with aforementioned expression, average bond distortion converges to the zero value, while number of hole doping increases. Interestingly, all types of bonds has similar behavior and bond distortion gets smaller and smaller.

**Conclusion**

We shown here the manifestation of the relaxation mechanism of a very small flake. The bond length changes of the graphene in nm size was considered as the number of holes increased and the result was compared with and without doping holes. Doping proceeds in three different categories: light, heavy and the most heavy doping. A pure flake is known as no hole doping, and also shows relaxation in nm size. Only 11 bonds out of 26 stretch out, and all vertical bonds also shrink. The stretching bonds number increases as hole doping increases.

Moreover, the most stable bonds are situated in the middle part of the flake, and the most unstable bonds usually are situated on the edge of the graphene especially for the light doping process. In all doped flakes, one does not estimate all relaxation in the bonds, but the flake's width and length become larger as doping holes increase. On the other hand, average bond distortion exhibits a non ordinary expectation. Thus our result suggest that one should consider both the distortion  and average distortion of all bonds in order to evaluate the bond distortion phenomenon of the graphene flake. There is no need to continue with more hole doping since the bond relaxation shows similar behavior, and no sharp changes appear after ten holes doping. In short, doping holes as well as doping electrons play crucial roles on the bond structure of a graphene flake in nm size.

## Figure Captions

**Fig. 1:** A graphene flake. Labeled numbers represent carbon atoms. Circled atoms that have unstable bonds and do not contribute to the calculation.

**Fig. 2(a):** Light hole doping. Relaxation in nm size against the bond between related carbon atoms. Solid, dotted and dashed lines represent the pure graphene flake, one hole doped graphene and two holes doped graphene, respectively.

**Fig. 2(b):** Signs (+) and (-) correspond to stretching and shrinking in the bond length. Graphs reflect from top to bottom: intrinsic, with one hole doped and two holes doped graphenes.

**Fig. 3(a):** Same as **Fig. 2(a)** but solid, dotted and dashed lines represent the pure graphene flake, three holes doped graphene and four holes doped graphene, respectively.

**Fig. 3(b):** Same as **Fig. 2(b)** but with three holes doped and four holes doped graphenes.

**Fig. 4(a):** Heavy hole doping. Same as **Fig. 3(a)** but solid, dotted and dashed lines represent the pure graphene flake, five holes doped graphene and six holes doped graphene, respectively.

**Fig. 4(b):** Same as **Fig. 3(b)** but with five holes doped and six holes doped graphenes.

**Fig. 5(a):** Same as **Fig. 4(a)** but solid, dotted and dashed lines represent the pure graphene flake, seven holes doped graphene and eight holes doped graphene, respectively.

**Fig. 5(b):** Same as **Fig. 4(b)** but with seven holes doped and eight holes doped graphenes.

**Fig. 6(a):** The most heavy hole doping. Same as **Fig. 5(a)** but solid, dotted and dashed lines represent the pure graphene flake, nine holes doped graphene and ten holes doped graphene, respectively.

**Fig. 6(b):** Same as **Fig. 5(b)** but with nine holes doped and ten holes doped graphene.

**Fig. 7:** Average bond distortion against number of the hole doping. Solid, dotted and dashed lines represent the tilt (horizontal), right (vertical) and tilt+right (all) bonds, respectively.

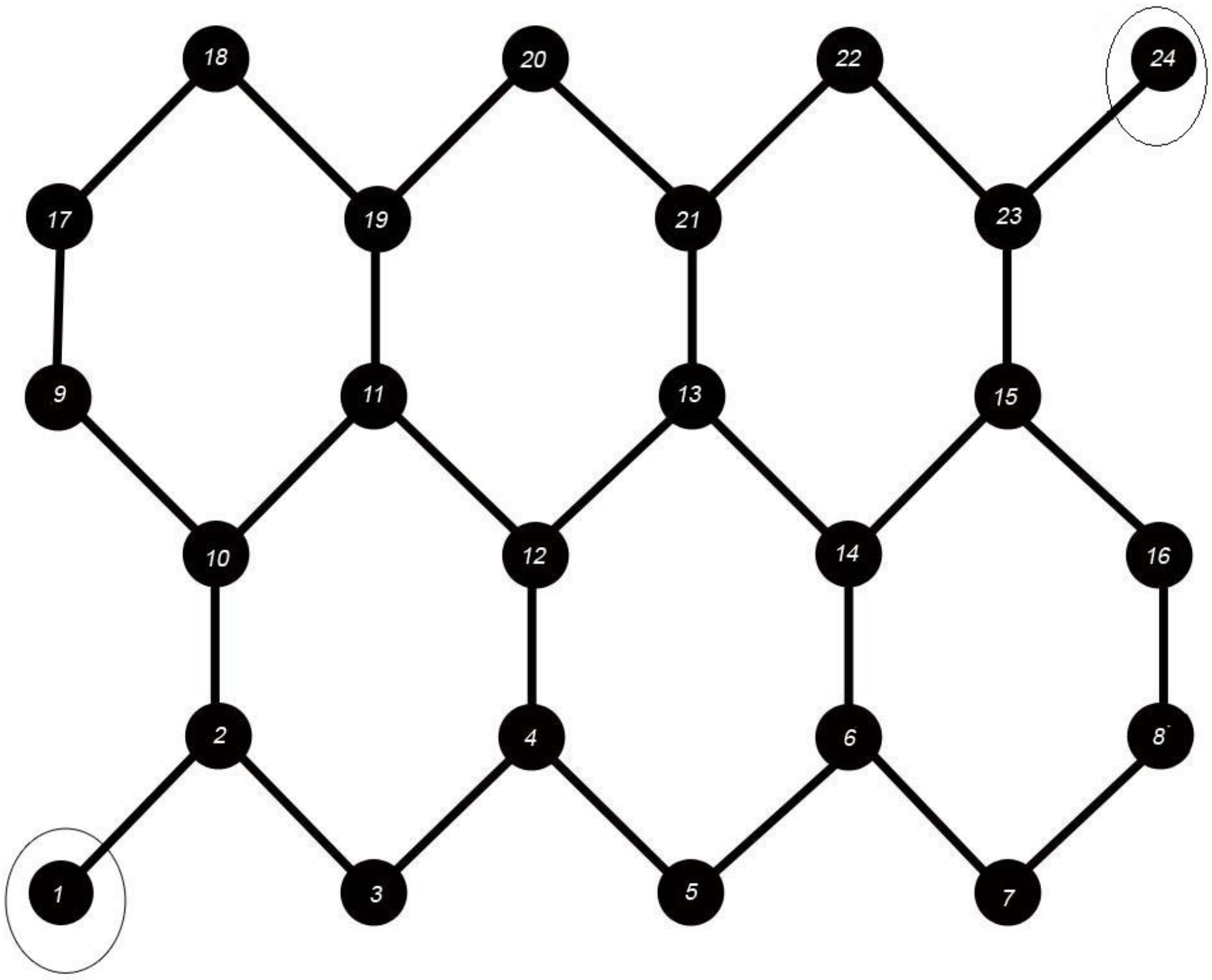

Fig. 1

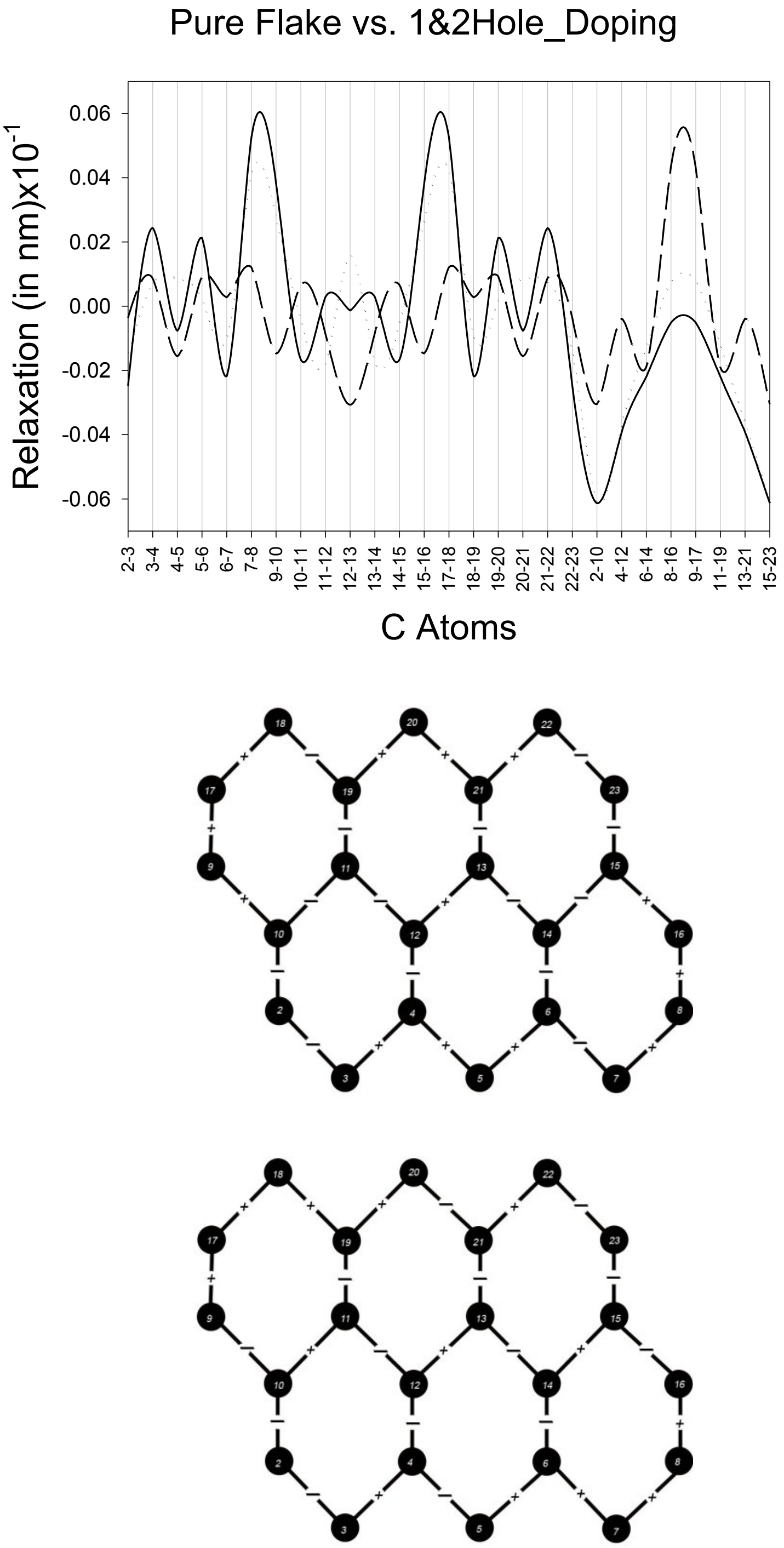
Pure Flake vs. 1&2Hole_Doping
Relaxation (in nm)x10$^{-1}$
0.06
0.04
0.02
0.00
-0.02
-0.04
-0.06
2-3
3-4
4-5
5-6
6-7
7-8
9-10
10-11
11-12
12-13
13-14
14-15
15-16
17-18
18-19
19-20
20-21
21-22
22-23
2-10
4-12
6-14
8-16
9-17
11-19
13-21
15-23
C Atoms

Fig. 2

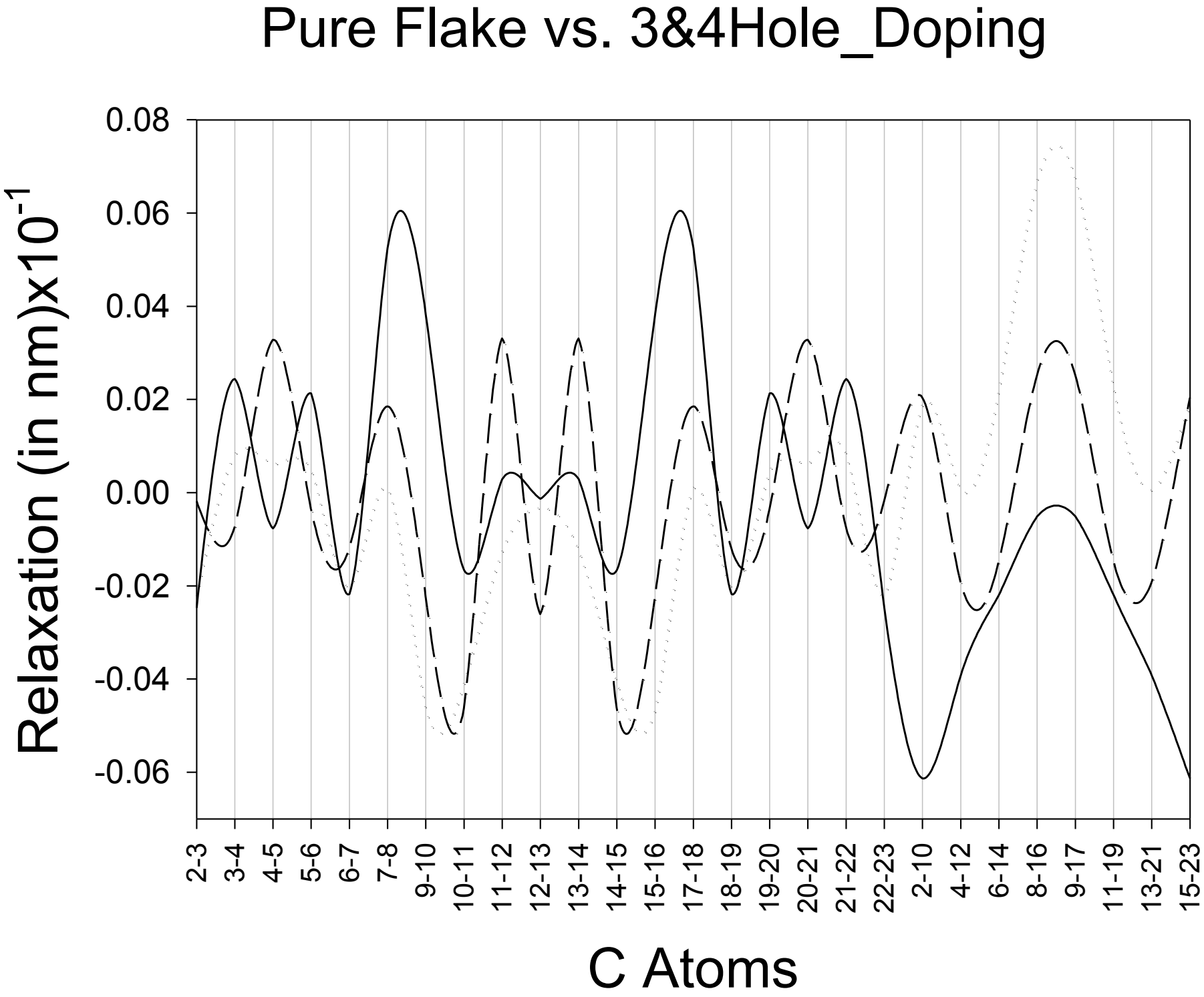
Pure Flake vs. 3&4Hole_Doping
Relaxation (in nm)x10$^{-1}$
0.08
0.06
0.04
0.02
0.00
-0.02
-0.04
-0.06
2-3
3-4
4-5
5-6
6-7
7-8
9-10
10-11
11-12
12-13
13-14
14-15
15-16
17-18
18-19
19-20
20-21
21-22
22-23
2-10
4-12
6-14
8-16
9-17
11-19
13-21
15-23
C Atoms

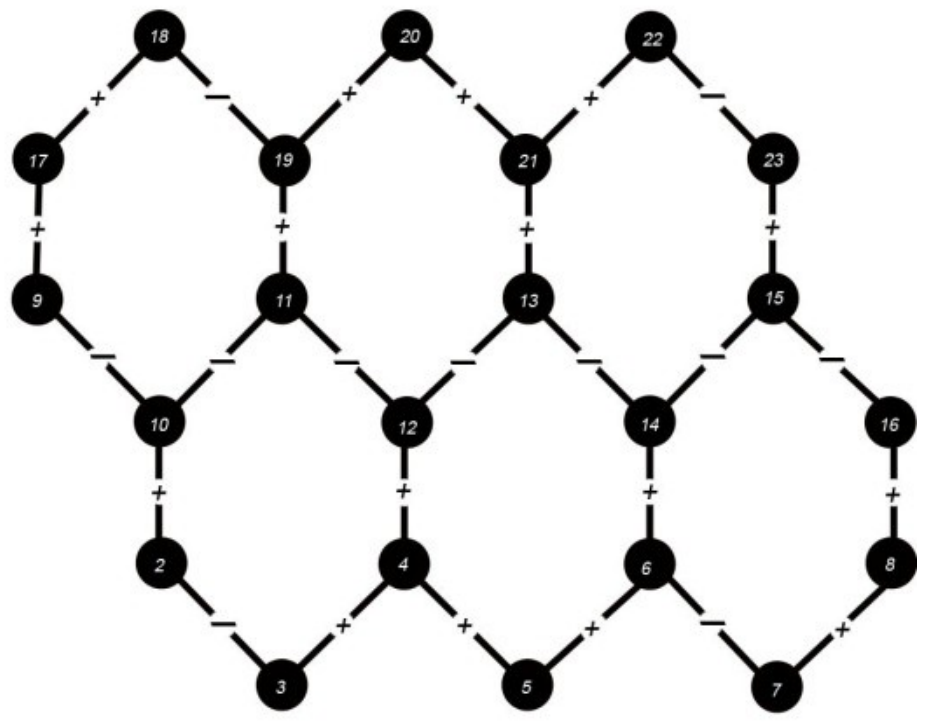

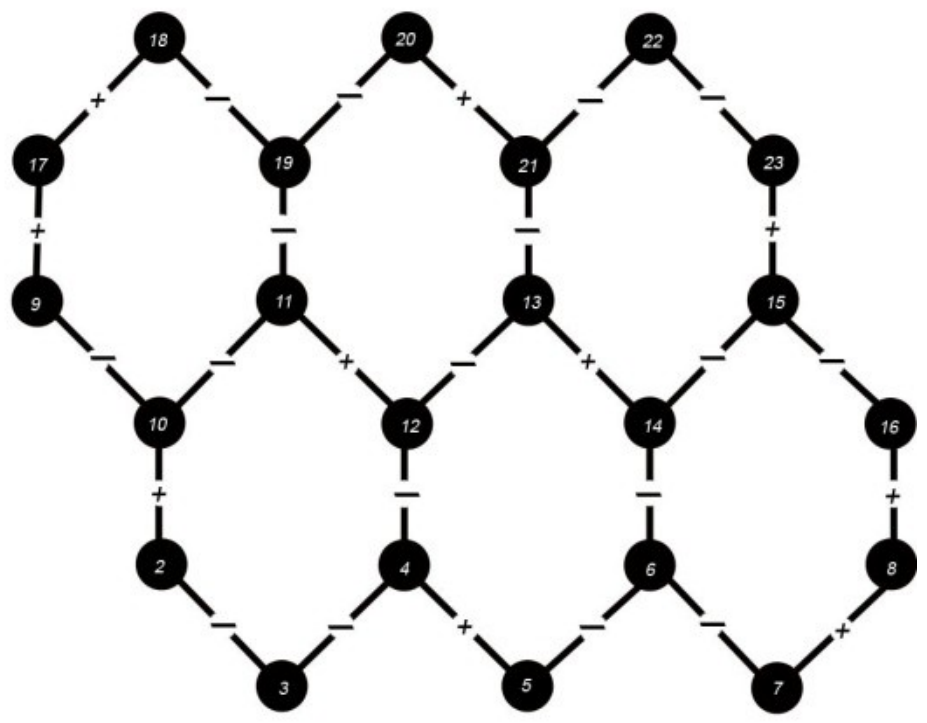

Fig. 3

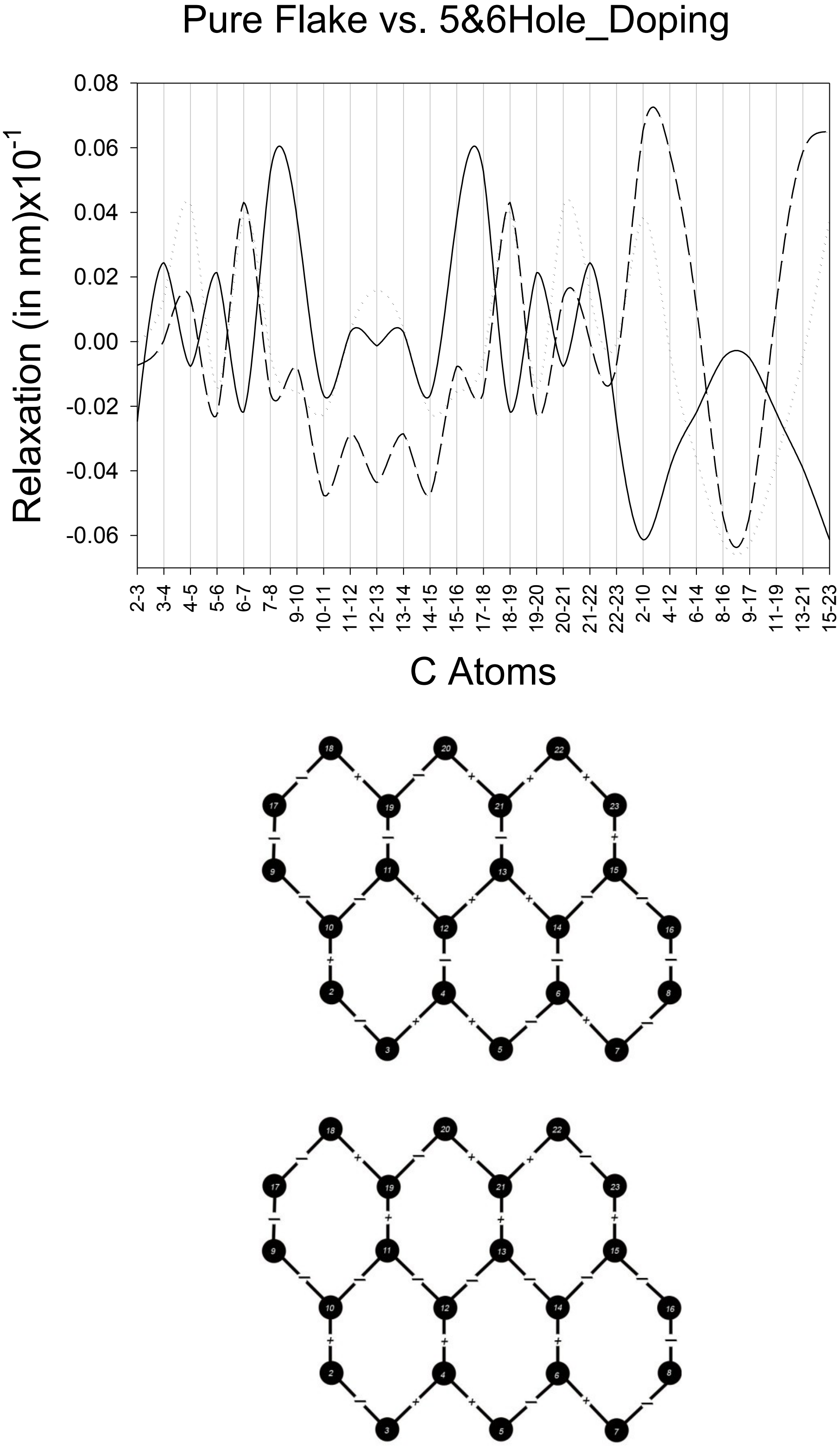

Pure Flake vs. 5&6Hole_Doping
Relaxation (in nm)x10$^{-1}$
0.08
0.06
0.04
0.02
0.00
-0.02
-0.04
-0.06
2-3
3-4
4-5
5-6
6-7
7-8
9-10
10-11
11-12
12-13
13-14
14-15
15-16
17-18
18-19
19-20
20-21
21-22
22-23
2-10
4-12
6-14
8-16
9-17
11-19
13-21
15-23
C Atoms

Fig. 4

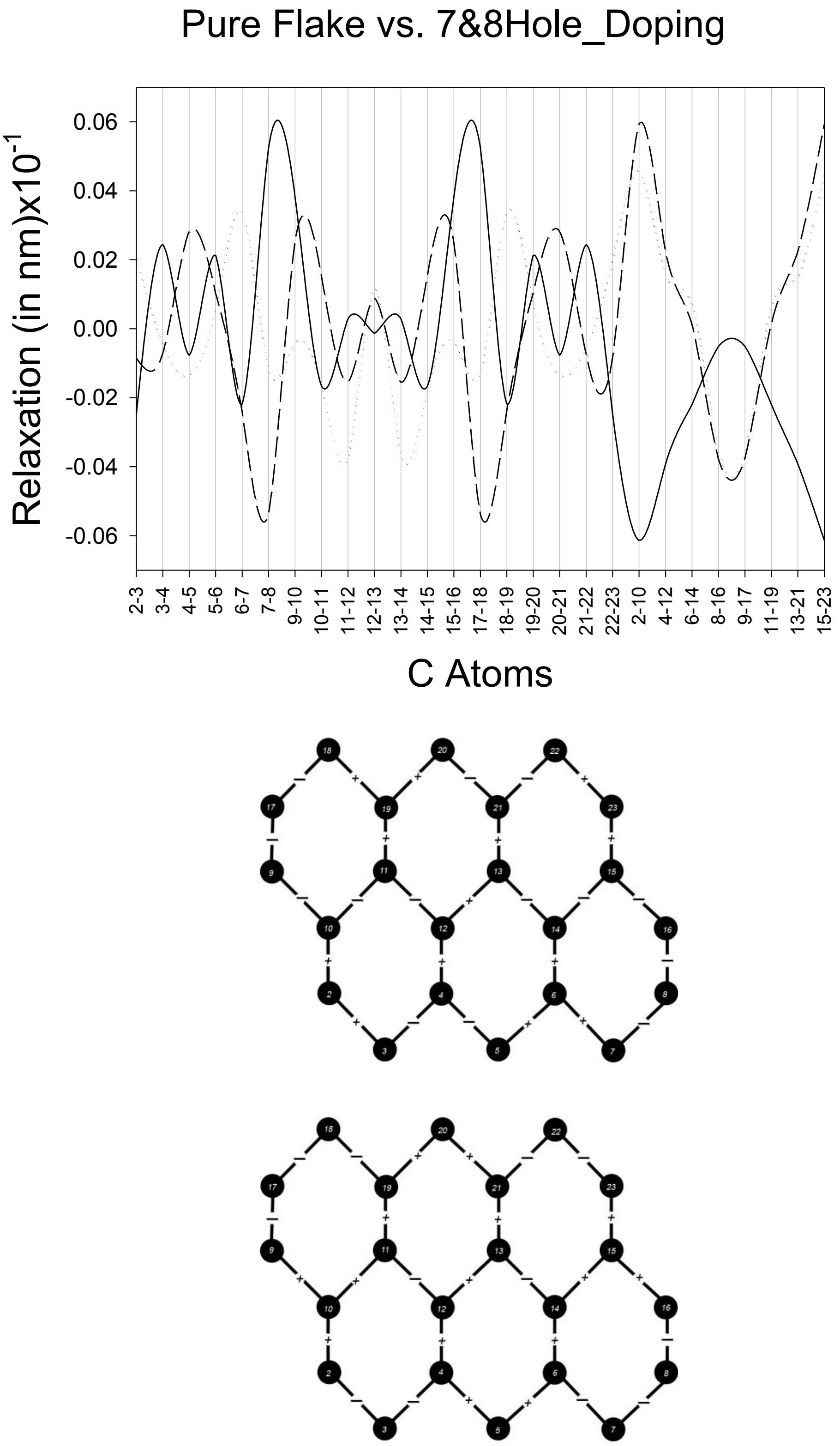
Pure Flake vs. 7&8Hole_Doping
Relaxation (in nm)x10$^{-1}$
0.06
0.04
0.02
0.00
-0.02
-0.04
-0.06
2-3
3-4
4-5
5-6
6-7
7-8
9-10
10-11
11-12
12-13
13-14
14-15
15-16
17-18
18-19
19-20
20-21
21-22
22-23
2-10
4-12
6-14
8-16
9-17
11-19
13-21
15-23
C Atoms

Fig. 5

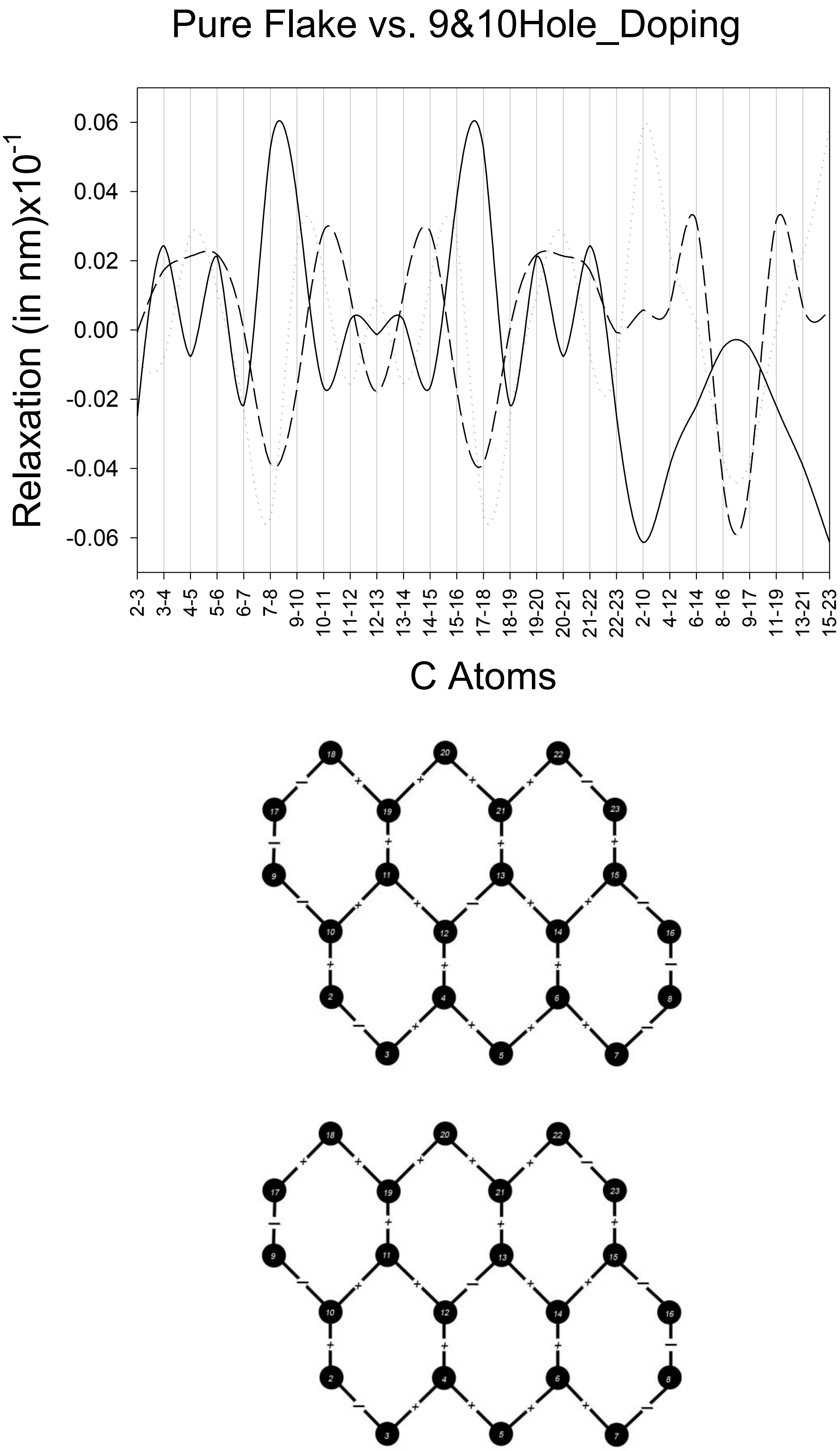
Pure Flake vs. 9&10Hole_Doping
Relaxation (in nm)x10^-1
0.06
0.04
0.02
0.00
-0.02
-0.04
-0.06
2-3
3-4
4-5
5-6
6-7
7-8
9-10
10-11
11-12
12-13
13-14
14-15
15-16
17-18
18-19
19-20
20-21
21-22
22-23
2-10
4-12
6-14
8-16
9-17
11-19
13-21
15-23
C Atoms

Fig. 6